# Sequence-dependent sensitivity explains the accuracy of decisions when cues are separated with a gap

Maryam Tohidi-Moghaddam[a, b], Sajjad Zabbah[b], Farzaneh Olianezhad[c, a], Reza Ebrahimpour[a, b]∗

[a] Faculty of Computer Engineering, Shahid Rajaee Teacher Training University, P.O. Box: 16785-136, Tehran, Iran

[b] School of Cognitive Sciences, Institute for Research in Fundamental Sciences (IPM), P.O. Box: 19395-5746, Tehran, Iran

[c] Faculty of Electrical Engineering, Shahid Rajaee Teacher Training University, P.O. Box: 16785-163, Tehran, Iran

**Abstract**

Most decisions require information gathering from a stimulus presented with different gaps. Indeed, the brain process of this integration is rarely ambiguous. Recently, it has been claimed that humans can optimally integrate the information of two discrete pulses independent of the temporal gap between them. Interestingly, subjects' performance on such a task, with two discrete pulses, is superior to what a perfect accumulator can predict. Although numerous neuronal and descriptive models have been proposed to explain the mechanism of perceptual decision-making, none can explain human behavior on this two-pulse task. In order to investigate the mechanism of decision-making on the noted tasks, a set of modified drift-diffusion models based on different hypotheses were used. Model comparisons clarified that, in a sequence of information arriving at different times, the accumulated information of earlier evidence affects the process of information accumulation of later evidence. It was shown that the rate of information extraction depends on whether the pulse is the first or the second one. The proposed model can also explain the stronger effect of the second pulse as shown by Kiani et al. (2013).

∗ Correspondence to: R. Ebrahimpour, Department of Computer Engineering, Shahid Rajaee Teacher Training University, P.O. Box: 16785-163, Tel/Fax Number: +982122970117, Tehran, Iran. E-mail addresses: m.tohidi@ipm.ir (M. Tohidi-Moghaddam), s.zabbah@ipm.ir (S. Zabbah), f.olianezhad@ipm.ir (F. Olianezhad), rebrahimpour@srttu.edu (R. Ebrahimpour).

## 1. Introduction

Struggling with an extensive range of decisions with various degrees of complexity and type is an inseparable part of life, from voting to a competent president by integrating various discrete information over a period of time to simple immediate decisions such as picking up a pen. Recent studies on perceptual decisions in primates have investigated the neurobiological mechanisms underlying simple decisions such as a binary choice between two likely stimuli (*Britten et al., 1996; Heekeren et al., 2004; Horwitz et al., 2004; Kiani et al., 2008;Latimer et al., 2015; Newsome et al., 1989; Roitman & Shadlen, 2002; Schall, 2003*). Responses of neurons in the lateral intra-parietal area (LIP) gradually increase to reach a specific level of firing rate when a monkey wants to make a decision about the motion direction in a stimulus with its eye movement (*Cook et al., 2002, Gold & Shadlen, 2001, 2007; Hanks et al., 2014; Heitz & Schall, 2012; Huk & shadlen, 2005; Kiani & Shadlen, 2009; Kiani et al., 2008; Philiastides & Sajda, 2005*). Therefore, it has been hypothesized that a variety of perceptual and mnemonic decisions in primates can be explained by a ''bounded evidence accumulation'' framework (*Kira et al., 2015; Philiastides & Sajda, 2005; Ratcliff & McKoon, 2008; Ratcliff & Tuerlinckx, 2002; Usher & McClelland, 2001*). A well-known class of models, drift diffusion model (DDM) and its extensions, has been proposed to explain this process based on the integration of noisy evidence in favor of different alternatives to reach a decision bound (*Gold & Shadlen, 2007; Link, 1992; Ratcliff & McKoon, 2008*). Different extensions of the DDM can interpret the reaction time, accuracy, confidence, speed-accuracy trade-off, and some other aspects of decision-making (*Heitz & Schall, 2012; Hanks et al., 2014; Kiani et al., 2014; Smith & Vickers, 1988; Usher & McClelland, 2001; Wang, 2002*).

Indeed, the vast majority of decisions involve a complicated procedure which is the accumulation of evidence from separated cues over time. Along the way, the brain needs to evaluate and integrate various pieces of information arriving at different times (*Kiani et al., 2013; Kira et al., 2015; Yang & Shadlen, 2007*). Recently, it has been demonstrated that the brain can optimally accumulate evidence from a set of noisy observations and attach more weight to the more decisive evidence (*Kira et al., 2015; Yang & Shadlen, 2007*). It has also been manifested that humans can gather information from two discrete cues separated by a gap of time, independent of the length of gaps. More importantly, Kiani et al. indicated that the latest evidence has far more influence on decisions (*Kiani et al., 2013*). However, it still remains obscure which mechanism can explain the process of decision-making in the presence of two pulses separated by time gaps.

The aim of this study is to investigate the mechanism of decision-making a two-pulse experiment suggested by Kiani et al. (2013). A behavioral experiment including trials with one (single-pulse) and two (double-pulse) pulses of random-dot motion (similar to *Kiani et al., 2013*) was implemented. First, subjects' accuracy improved in double-pulse trials more than what had been expected from single-pulse trials, and accuracy was independent of the interim period between pulses. Moreover, the second pulse had a larger effect on subjects' performance. These results are the replication of what Kiani et al. (2013) had previously clarified.

Using a well-known decision-making model, DDM, the plausible mechanism underlying this behavior was explored (*Gold & Shadlen, 2007; Mazurek et al., 2003; Shadlen et al., 2006*). The DDM contains three basic parameters of starting point, drift rate, and width of the interval between decision thresholds (*Bogacz et al., 2006; Ratcliff, 1978; Ratcliff & Rouder, 1998; Ratcliff & Tuerlinckx, 2002; Ratcliff et al., 2016*). According to the parameters of the model, three different hypotheses were examined. Results showed that the process of information accumulation differs across double-pulse and single-pulse trials. The accumulated information in the first pulse plays the role of a starting point for the accumulation process in the second pulse. More importantly, the rate of accumulation, not the bound, depends on the sequence of pulse presentation. Interestingly, the proposed model can well explain the sequence-dependent behavior in trials with different pulse strengths, even when the parameters are fitted using different sets of trials with the same pulses strength.

## 2. Material and Method

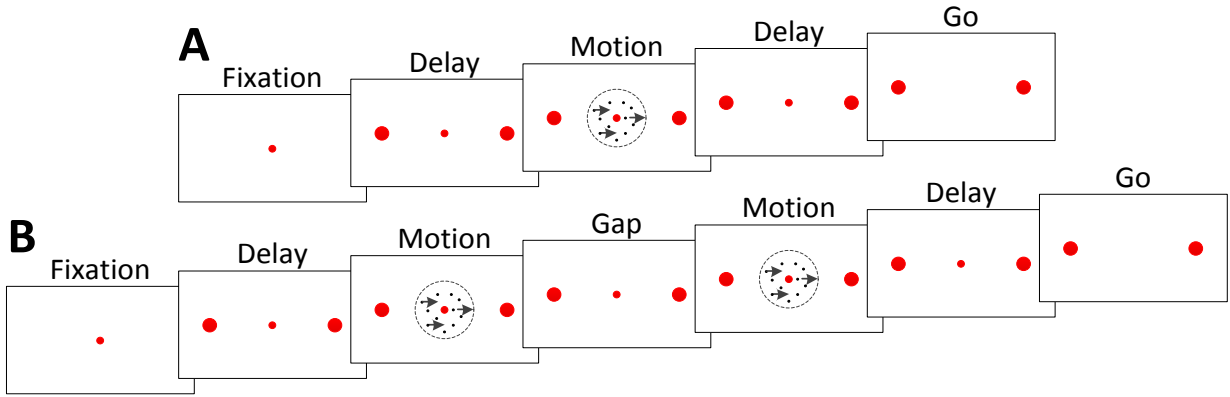

Figure 1. Behavioral task paradigm. All trials started with a red fixation point on the center of the screen. Whenever the subject was ready and pressed a key, two red target points appeared on two sides of the fixation point for a random delay. Then, motion presentation started for one (A) or two (B) pulses. After stimulus presentation, fixation and two red target points were shown for a random delay. Finally, subjects received the Go cue and, after indicating the choice, received feedback of their answers.

### 2.1. Subjects

Four human subjects, two females and two males aging 24 to 31 were tested in this experiment. Two subjects (1 female and 1 male) were naive to the purpose of the experiment, and all subjects had normal or corrected-to-normal vision. Subjects signed an informed consent form before participating in the experiment. The Ethics Committee of Iran University of Medical Sciences approved the experimental protocol.

Before the main experiment, subjects were extensively trained to achieve a high level of accuracy and a predefined reaction time. In the main experiment, each subject completed 11–15 blocks containing 300 trials.

### 2.2. Stimulus

Subjects were asked to discriminate the direction of motion stimuli consisting of three independent sets of dots displayed on consecutive video frames (dot density: 16.7 dots/deg$^2$/s). The size of dots was 2×2 pixels (side: 0.088°) (*Roitman & Shadlen, 2002;* Shadlen & Newsome, 2001). Each set of white dots was shown on a black background for one frame (13.3 millisecond (ms)) and was updated three frames later (after 40 ms). Therefore, the positions of the dots in frames one, two, and three were respectively updated in frames four, five, and six, and so on. To update, each dot was either moved coherently along the motion direction (right or left) at the velocity of 6 deg/s or relocated randomly. Task difficulty was controlled through the coherence of motion which is the probability that each dot would be moved toward the motion direction.

### 2.3. Procedure

All experiments were conducted in a dimly lit room. Subjects were seated in front of a 19" LG CRT display monitor (screen resolution: 800×600, refresh rate: 75 Hz) at the distance of 57 cm. An adjustable chin rest and a forehead bar were used to ensure the stability of subjects' heads. MATLAB's Psychtoolbox (Brainard & Vision, 1997; Pelli, 1997) was employed to display the stimuli. In the following, the procedure of trials is described.

1- **Fixation**: A red fixation point with a 0.3° diameter at the center of the screen was on during the trial to maintain subjects' attention. Each trial started with the appearance of the red fixation point. Then, subjects were asked to gaze at the fixation point and maintain their gaze within the trial. Upon pressing the space key, the second step would start.

2- **Target on**: Two red targets with a 0.5° diameter were presented on the right and left sides of the fixation point at a distance of 10° to cue subjects for possible direction of motion. After a random delay (200–500 ms, truncated exponential), the motion stimuli were presented.

3- **Motion**: The motion stimulus was displayed within a 5° circular aperture at the center of the monitor. As mentioned before, two types of stimulus presentation, single-pulse and double-pulse trials, were utilized as below:
    - Single-pulse: Motion coherence was randomly selected from these values: 0%, 3.2%, 6.4%, 12.8%, 25.6%, and 51.2%. After 120 ms of stimulus presentation, the next step was followed (Fig. 1, A).
    - Double-pulse: In each pulse, motion coherence was randomly selected from these values: 3.2%, 6.4%, and 12.8%. However, both pulses had the same net direction of motion, and subjects were informed. Two pulses were separated by a gap from

these values: 0, 120, 360, or 1080. In total, there were 9×4 types of trials. Each pulse was presented for 120 ms and, after the second pulse, the next step was followed (Fig. 1, B).

4- **Delay**: As in the second step, two targets and a fixation point were shown for a random delay (400–1000 ms, truncated exponential).

5- **Go**: After the random delay, the fixation point extinguished and subjects were instructed to report their decision within 1 sec after the Go cue by pressing two specific keys for right and left choices. After that, an auditory feedback with two frequencies (low frequency for correct and high frequency for error) beeped for correct and error responses. More importantly, on trials with 0% motion coherence, the frequency of auditory feedback was randomly selected.

Subjects completed one or two blocks a day, each block including 300 trials. In total, 42 (6+36) types of trials were randomly interleaved in each block. Subjects were asked to perform each trial as accurately as possible. The probability correct of each block was displayed at the end of the block as a feedback to the subjects.

### 2.4. Data Analysis

To estimate the effect of parameters on the binary choices, different logistic regression models were applied. In the following regression models, Logit[P] indicates log *(P/1-P)* and $\beta_i$ denotes fitted coefficients.

To approximate the performance of single-pulse trials, the first regression model was fitted as below:

$$\text{Logit}[P_{correct}] = \beta_0 + \beta_1 C \quad [1]$$

where C is the coherency level of single-pulse trials.

To evaluate whether the probability correct is affected by the interim period or not, the following regression analysis was utilized:

$$\text{Logit}[P_{correct}] = \beta_0 + \beta_1 C_1 + \beta_2 C_2 + \beta_3 T + \beta_4 C_1 T + \beta_5 C_2 T \quad [2]$$

where $C_1$ and $C_2$ are the motion coherence of the first and second pulse, respectively, and T is the interim period. The null hypothesis is the lack of relationship between the probability correct and the interim period ($H_0$: $\beta_{3-5}$ =0).

In the next step, a regression model was fitted to compare the expected accuracy of a perfect integrator to the observed performance on discrete trials:

$$\text{Logit}[P_{correct}] = \text{Logit}[P_e] + \beta \quad [3]$$

where $P_e$ is the expected probability correct resulting from a perfect accumulator (see *Kiani et al., 2013*, Eqs. 4 and 5). A positive value of β indicates that the observed accuracy is higher than the expected one.

The following regression model was used to test the effect of pulse sequence on the probability correct:

$$\text{Logit}[P_{correct}] = \beta_0 + \beta_1[C_1 + C_2] + \beta_2[C_1 - C_2] \quad [4]$$

where $C_1$ and $C_2$ are the motion coherence of the first and second pulse, respectively, and $\beta_2$ represents the change in probability correct from the weak-strong pulse ($C_1 < C_2$) to the strong-weak one ($C_1 > C_2$). The null hypothesis is that the probability correct does not depend on the pulse sequence ($H_0$: $\beta_2$ = 0).

### 2.5. Modeling

To explain the plausible mechanism for decisions with two discrete pulses, the DDM (Eq. 5) (*Ratcliff, 1978; Ratcliff & McKoon, 2008*) implemented by Voss et al. was applied in a computationally efficient and accessible program called fast-dm (*Lerche & Voss, 2017; Voss & Voss, 2007; Voss et al., 2004, 2013*).

$$dv = Vdt + \xi\sqrt{dt} \text{ where } v(0) = z \quad [5]$$

This model is described via an accumulation-to-bound mechanism that starts with gathering noisy sensory evidence (explained by drift rate (v)) from a starting point (z) and terminates upon crossing the pre-specified response criterion (a). These three basic parameters existing in early versions of the diffusion model are compatible with both neurophysiological and behavioral data. The refined version of DDM was designated by seven parameters grouped into the following three classes: (Ⅰ) the basic parameters of the decision process (decision threshold a, mean starting point z, and mean drift rate v), (Ⅱ) the non-decision process parameter (non-decision time $t_{ND}$), and (Ⅲ) the changeability across-trial parameters (changeability in stimulus quality $\eta$, changeability in starting point sz, and starting point in non-decision time $st_{ND}$) (*Ratcliff, 1978; Ratcliff & McKoon, 2008; Ratcliff & Tuerlinckx, 2002*).

Fast-dm estimated the DDM's parameters using the partial differential equation (PDE) method (*Voss & Voss, 2007*) which had swift calculations to compute the cumulative distribution function (CDF) and the Kolmogorov-Smirnov statistic (*Voss et al., 2004*). The diffusion model can describe the perceptual decision process in the brain and is actually a well-entrenched model in this field (*Ratcliff, 1978; Ratcliff & McKoon, 2008*).

In this study and based on our hypothesis, three modified models were employed and compared using the Bayesian information criterion (BIC) (*Kass & Wasserman, 1995; Liddle, 2007; Smith & Spiegelhalter, 1980*).

## 3. Results

Four human subjects were asked to make a decision about the net direction of motion in trials in which a 120ms motion pulse was displayed once or twice. Single-pulse and double-pulse trials were randomly presented. In double-pulse trials, both pulses had the same direction, and subjects were informed about this consistency. The net direction, motion strength, and interim period were random from trial to trial. Here, we first demonstrated that our data can reproduce the results reported by Kiani et al. (2013), and then fitted different DDMs to shed light on the mechanism of decision-making under discrete conditions.

Since 86% of trials were double-pulses, it may be hypothesized that the first pulse would be ignored and the subjects decided merely based on the second pulse. However, the psychometric function of the single-pulse trials (Fig. 2) showed that the subjects accurately used the information of the first pulse. As expected from previous studies, the probability of correct increased as a function of motion strength (*Kiani et al. 2013*), supporting the claim that subjects did not ignore the first pulse.

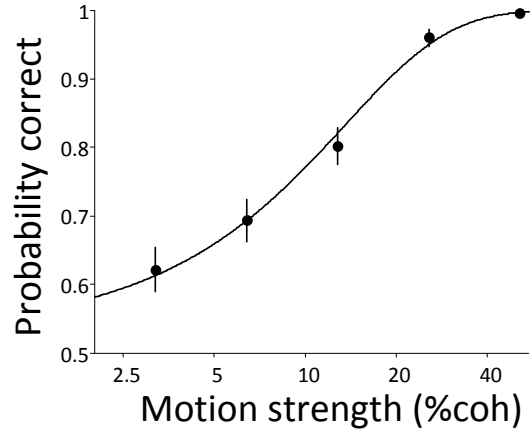

Figure 2. Psychometric function of the single-pulse trials. Each data point represents pooled data across all trials of all subjects. The solid curve is the fit of a regression model (Eq. 1) to the data. Error bars declare standard error of mean (SEM).

### 3.1. Achieving the same accuracy levels independent of the interim period

Consistent with previous results, accuracy was invariant to the interim period between pulses in double-pulse trials for both similar (Fig. 3A, Eq. 2, $\beta_3=0.31\pm0.02$, p=0.19; $\beta_4=-0.05\pm0.03$ p=0.15) and different (Fig. 3B, Eq. 2, $\beta_3=-0.11\pm0.29$, p=0.70; $\beta_4=0.01\pm0.02$ p=0.57; $\beta_5=-0.002\pm0.02$, p=0.92) pulses. Accordingly, this finding demonstrated that the information of the first pulse was not lost but maintained throughout the interim period. Furthermore, the probability correct in weak-strong double-pulse trials was enhanced to above the expected performance of a perfect accumulator (Eq. 3, $\beta=0.10\pm0.04$, p=0.018) which, at least, provides evidence for the use of the information of both pulses. However, the exceeded performance raised the hypothesis that the process of decision-making in double-pulse trials may be different from simply doubling the information of one pulse.

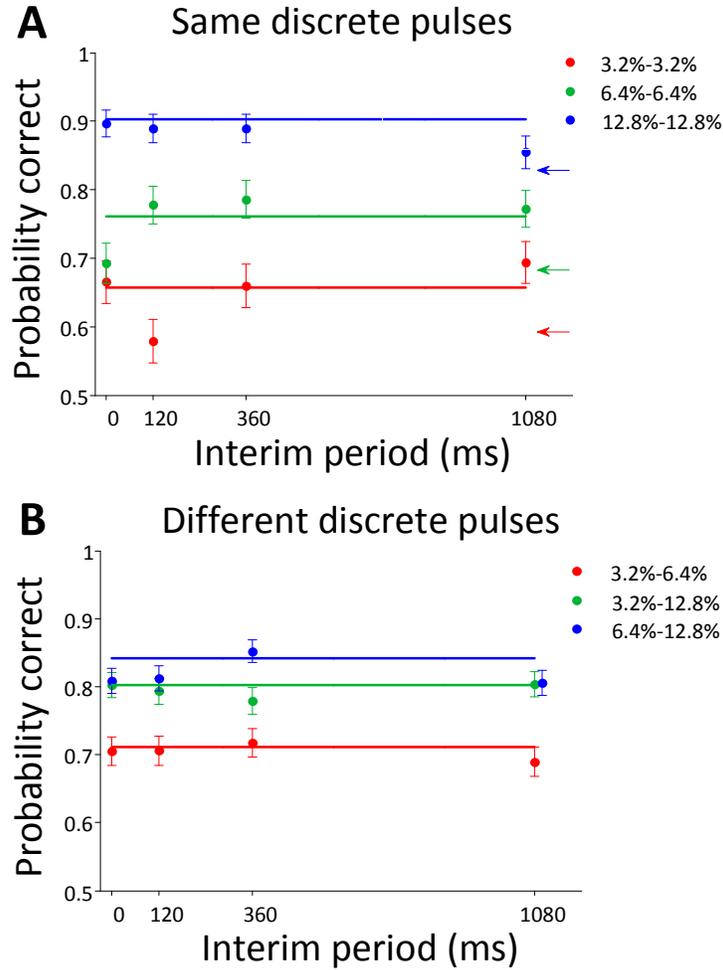

Figure 3. Probability correct of double-pulse trials, A) when pulses were similar; arrows show the performance value derived from the fitted regression model (Eq. 1) to the relevant single-pulse trials; solid lines indicate the performance of the perfect accumulator (Eq. 3) according to the single-pulse accuracy; B) when pulses were different; solid lines indicate the performance of the perfect accumulator (Eq. 3) according to the single-pulse accuracy; each data point shows pooled date from different double-pulse trials represented by legend and its reverse sequence (e.g., 3.2%-12.8%, 12.8%, and 3.2%). Error bars declare SEM.

### 3.2. What mechanism can explain the sequence-dependent performance of subjects in double-pulse trials?

Our data also replicated what Kiani et al. (2013) had demonstrated about the sequence of pulses. As stated above, subjects performed better than expected in weak-strong double-pulse trials and, as illustrated in Fig. 4, accuracy increased in double-pulse trials when the motion strength of the second pulse was stronger than that of the first one (Eq. 4, $\beta_2=0.03\pm0.005$,

p=8.60×10$^{-8}$). In this part of the Results section, we propose a plausible mechanism for the observed sequence-dependent performance which caused the over-expected accuracy in double-pulse trials.

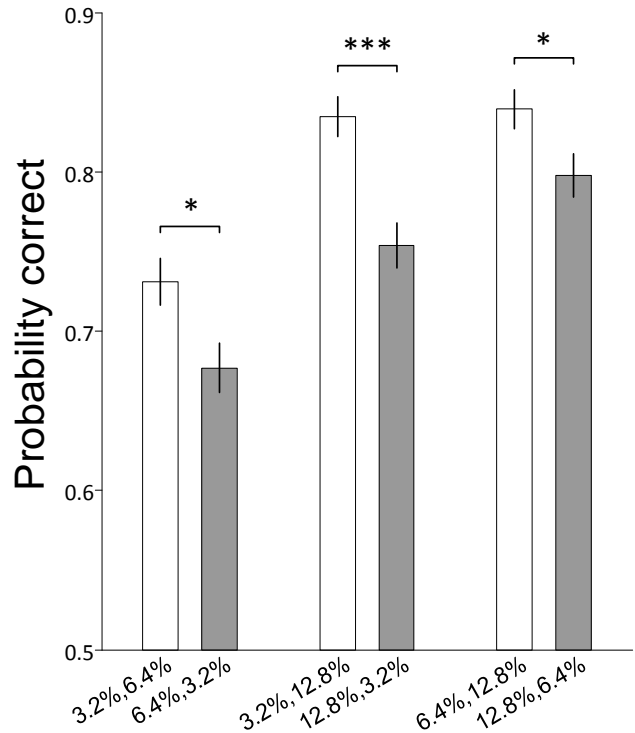

Figure 4. Probability correct dependency on the sequence of motion pulses in different discrete trials. Probability correct was calculated by pooling data across all interim periods. Error bars declare SEM. *p<0.05 and ***p<1E-3

Assuming the accumulation to bound model as a true model of the decision process, we used the DDM to investigate three plausible hypotheses for the mechanism of the decision process in double-pulse trials: 1) Decision making in double-pulse trials has the same mechanism as single-pulse trials such that information accumulation in the first pulse will be continued in the second pulse until crossing a bound. It means that the rate of accumulation (drift rate) in the diffusion model depends on the summation of coherency of two pulses; 2) the first pulse affects the rate of information accumulation in the second pulse; and 3) the first pulse affects the decision threshold or the bound of information accumulation in the second pulse.

In the following section, the result of examining these three hypotheses are presented.

### 3.3. Model Fitting

We used the DDM to test our hypotheses and examine the mechanism of decision-making in double-pulse trials. In so doing, based on our alternative hypotheses, we ran and compared three variations of the DDM to explore how and which parameters were influenced by discretizing information.

In the first model, we assumed that the information accumulation is similar in both single-pulse and double-pulse trials. It means that the interim period is not considerable, and the two pulses are considered as one pulse. To examine this assumption, we made the drift-rate ($v$) a function of the sum of the coherency of two double-pulses and named it model$_{sum}$. The second and third models were fitted to explore whether the process of decision-making in the second pulse is influenced by the accumulated information of the first pulse or not. To do so, in the second hypothesis, we assumed that the accumulated information of the first pulse can affect the rate of information accumulation in the second pulse. In order to check this hypothesis, we fitted model$_{starting\ point}$ and defined the starting point ($z$) and the drift-rate ($v$) of the DDM as a function of the motion strength of the first pulse and the second pulse, respectively. Assuming a perfect accumulator, the accumulated information at the end of the first pulse is the starting point of information accumulation in the second pulse. As a result, making the starting point of the accumulation in the second pulse dependent on the strength of the stimulus in the first pulse, we let the model consider different drift rates for the stimulus in the first and the second pulses. The third DDM model, model$_{bound}$, was based on the hypothesis that the first pulse can influence the decision bound of the information accumulation in the second pulse. Thus, we made the bound ($a$) of the DDM on the motion strength of the first pulse and the drift-rate ($v$) on the motion strength of the second pulse. The fitted parameters of each model are listed in Tables 1-3 (Standard errors (SEs) were calculated with 100 iterations bootstrap).

**Table 1.** Fitted parameters (mean±SE) of the first DDM (model$_{sum}$). $V_{3.2-6.4}$ indicates the fitted drift-rate of pooled data from the pulse sequence and its reverse order (e.g., 3.2– 6.4% and 6.4 –3.2%).

|   |   |
|---|---|
| $z$ | 0.5470±0.0159 |
| $a$ | 1.0903±0.0170 |
| $v_{3.2-6.4}$ | 0.7962±0.0956 |
| $v_{3.2-12.8}$ | 1.2964±0.1159 |
| $v_{6.4-12.8}$ | 1.3780±0.1045 |
| $t_{ND}$ | 0.7101±0.0056 |
| $st_{ND}$ | 0.3382±0.0133 |

**Table 2**. Fitted parameters (mean±SE) of the second DDM (model$_{starting\ point}$)

| | |
|---|---|
| $z\_3.2$ | 0.5206±0.0300 |
| $z\_6.4$ | 0.5425±0.0294 |
| $z\_12.8$ | 0.5772±0.0290 |
| $a$ | 1.0909±0.0238 |
| $v_{3.2}$ | 0.7604±0.1348 |
| $v_{6.4}$ | 1.0882±0.1430 |
| $v_{12.8}$ | 1.6308±0.1674 |
| $t_{ND}$ | 0.7181±0.116 |
| $st_{ND}$ | 0.3407±0.0150 |

**Table 3.** Fitted parameters (mean±SE) of the third DDM (model$_{boound}$)

| | |
|---|---|
| $z$ | 0.5379±0.0222 |
| $a\_3.2$ | 1.0829±0.0260 |
| $a\_6.4$ | 0.0773±0.0216 |
| $a\_12.8$ | 1.1160±0.0238 |
| $v_{3.2}$ | 0.8011±0.1162 |
| $v_{6.4}$ | 1.1178±0.1270 |
| $v_{12.8}$ | 1.6638±0.1452 |
| $t_{ND}$ | 0.7131±0.0082 |
| $st_{ND}$ | 0.3419±0.0146 |

Here, we utilized the BIC (*Kass & Wasserman, 1995; Liddle, 2007; Smith & Spiegelhalter, 1980*) to compare the models (Table 4). As shown in this table, the BIC of model $_{starting\ point}$ is smaller than that of the other models, indicating that this model can better explain the observed behavior of subjects in double-pulse trials.

Table 4. Model performance comparison via BIC and R2 metrics (mean±SD across subjects)

| Model | Total parameters | $R^2$ | BIC |
|---|---|---|---|
| Model$_{sum}$ | 15 | 0.8181±0.0550 | -29.7407±2.8348 |
| Model$_{starting\ point}$ | 14 | 0.8660±0.0573 | -35.1189±4.2153 |
| Model$_{bound}$ | 14 | 0.7470±0.0884 | -29.0620±3.3791 |

In the next step, we also investigated the difference among the dependent parameters for the best model (model$_{starting\ point}$). As mentioned in Table 2, the winner model produced three starting points and drift rates. Therefore, the significance of the differences among them was tested by the nonparametric bootstrap method (*Efron, 1992; Efron & Tibshirani, 1994*). These differences were highly significant ($p_{z3.2,z6.4} \ll 0.05$, $p_{z3.2,z12.8} \ll 0.05$, and $p_{z6.4,z12.8} \ll 0.05$).

### 3.4. Analyzing the parameters of the best model

Table 2 illustrates the parameters of the best model. Considering that each pulse lasts 120 ms, one can calculate the drift rate of the first pulse using Eq. 6:

$$\mu = \frac{(z-0.5)}{0.120} \quad [6]$$

where μ is the drift rate and z is the starting point for the second pulse or the accumulated information at the end of the first pulse. Moreover, 0.5 is the starting point of the first pulse. As a result, the drift rate of each stimulus at each pulse is presented in Table 5.

Table 5. Drift rate of the first and the second pulse

| | First pulse | Second pulse |
|---|---|---|
| $v_{3.2}$ | 0.172 | 0.6915 |
| $v_{6.4}$ | 0.3542 | 1.0125 |
| $v_{12.8}$ | 0.6433 | 1.5319 |

Values in Table 5 suggest that the best model assigns larger values to the drift rate of the second pulse compared to the first pulse.

### 3.5. Model_{starting point} can explain the sequence-dependent behavior of subjects

Fig. 5 demonstrates that model_{starting point} can mimic the same trend as behavioral data (Fig. 4). As reported by Kiani et al. (2013) and replicated here, subjects' accuracy in double-pulse trials is sequence-dependent. They rely more on the second pulse than on the first pulse to make their decisions. Here, we investigate whether the best model suggested by BIC and fitted on different-pulses trials (Fig. 5A) or same-pulses trials (Fig. 5B) can replicate the sequence-dependent behavior of subjects in trials with different pulses. This is notable to point out that, in Fig. 5B, the model is totally blind to the subject data in the trials with different pulses.

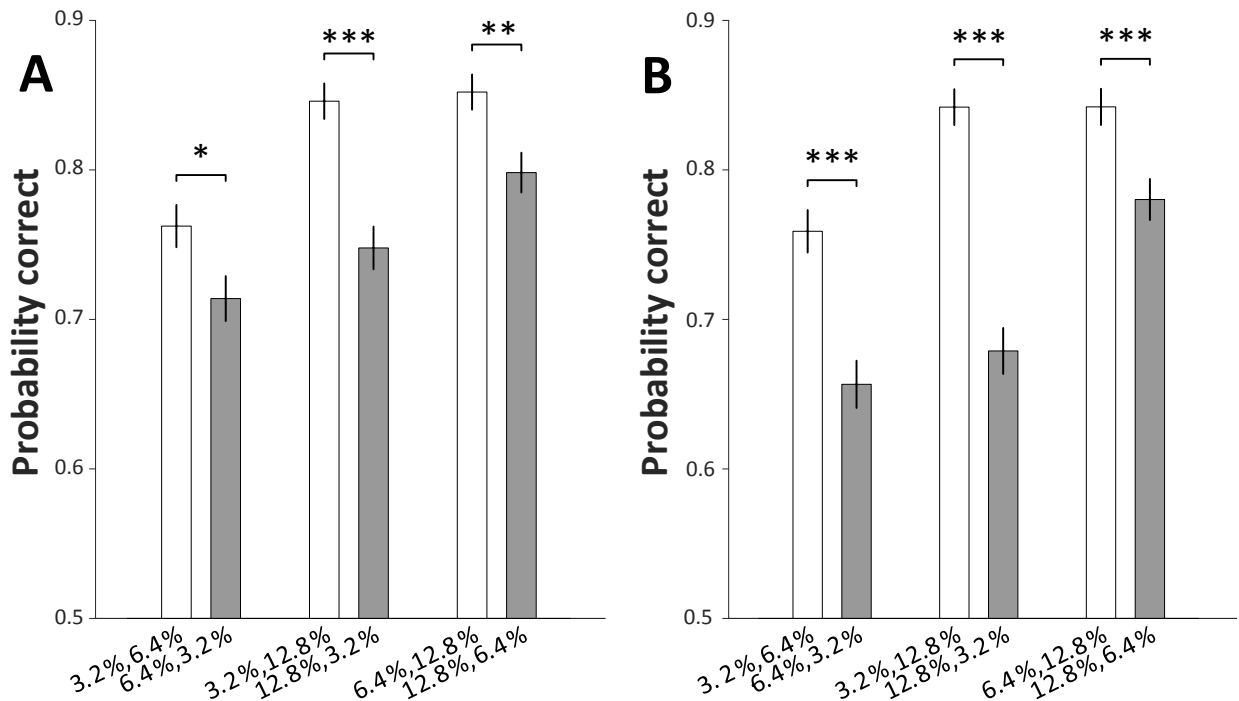

Figure 5. A) Simulated results of Fig. 4 with the fitted parameters of model_{starting point}. B) Simulated results of Fig. 4 when we fitted the parameters of our model in the same condition and simulated them in different conditions. Error bars denote SEM. ***p<1E-3.

## 4. Discussion

The decision-making process tends to integrate discrete pieces of information in favor of various options (*Kiani et al., 2013; Kira et al., 2015; Yang & Shadlen, 2007*). Kiani et al. (2013) showed that, in a fixed-duration direction discrimination task, the integration of direction cues is independent of the temporal gap between them. Moreover, they revealed that participants' accuracy is more affected by the last pieces of information than the primary ones and, more importantly, their accuracy in double-pulse trials is more than what is expected from their performance in single-pulse trials. Here, using a similar psychophysics task, in addition to the confirmation of previous results, we suggested a mechanism to explain participant behavior on such a task. Based on the DDM, we demonstrated that a sequence-dependent drift rate can better fit the behavioral results. The model also accounts for both the sequence dependence and the over-expected accuracy of participants. It is notable that Kiani et al. (2013) concluded that the accuracy of weak-strong and strong-weak trials is significantly higher than that of the perfect integrator fitted to the performance of single-pulse trials. However, this over-expected accuracy is reproduced in our results only in weak-strong trials.

Based on our results, the drift rate is not always a static parameter as assumed in previous studies. Based on Table 5, the rate of information extraction in the first pulse differs from that of the second pulse. This change in sensitivity also occurs in trials with similar pulse strength (same trials) in a way that the model fitted on the same trials can truly predict the sequence-dependent performance in different trials (Fig. 5B).

The change in the sensitivity of the second pulse can be explained by two hypotheses: 1) the temporal-dependency of sensitivity, according to which attention increases as the trial is near the end; and 2) the decision variable-dependency of sensitivity, according to which the sensitivity of the second pulse depends on the state of the accumulated evidence at the end of the first pulse. Nevertheless, since both time and accumulated evidence are correlated variables, distinguishing these hypotheses needs a more complex design of experiments.

## 5. Acknowledgments

This work was supported partially by Cognitive Sciences and Technologies Council under contract number 102 and Shahid Rajaee Teacher Training University under contract number 29602.